**Title:** Navigating the reporting guideline environment for computational pathology: A review

**Authors:** Clare McGenity[1,2], Darren Treanor[1,2,3,4]

**Affiliations:**
[1] University of Leeds, Leeds, UK
[2] Leeds Teaching Hospitals NHS Trust, Leeds, UK
[3] Department of Clinical Pathology and Department of Clinical and Experimental Medicine, Linköping University, Linköping, Sweden
[4] Centre for Medical Image Science and Visualization (CMIV), Linköping University, Linköping, Sweden

**Corresponding Author:** Dr Clare McGenity
**Email:** c.m.mcgenity@leeds.ac.uk

**Abstract**

The application of new artificial intelligence (AI) discoveries is transforming healthcare research. However, the standards of reporting are variable in this still evolving field, leading to potential research waste. The aim of this work is to highlight resources and reporting guidelines available to researchers working in computational pathology. The EQUATOR Network library of reporting guidelines and extensions was systematically searched up to August 2022 to identify applicable resources. Inclusion and exclusion criteria were used and guidance was screened for utility at different stages of research and for a range of study types. Items were compiled to create a summary for easy identification of useful resources and guidance. Over 70 published resources applicable to pathology AI research were identified. Guidelines were divided into key categories, reflecting current study types and target areas for AI research: Literature & Research Priorities, Discovery, Clinical Trial, Implementation and Post-Implementation & Guidelines. Guidelines useful at multiple stages of research and those currently in development were also highlighted. Summary tables with links to guidelines for these groups were developed, to assist those working in cancer AI research with complete reporting of research. Issues with replication and research waste are recognised problems in AI research. Reporting guidelines can be used as templates to ensure the essential information needed to replicate research is included within journal articles and abstracts. Reporting guidelines are available and useful for many study types, but greater awareness is needed to encourage researchers to utilise them and for journals to adopt them. This review and summary of resources highlights guidance to researchers, aiming to improve completeness of reporting.

**Introduction**

A typical patient is now projected to generate approximately one million gigabytes of medical data throughout the course of their lifetime.[1] The generation and availability of this electronic health data has presented both opportunities and challenges for artificial intelligence (AI) researchers. A wealth of uses have been described across the diverse healthcare technology landscape, with examples of AI applied to clinical imaging, electronic health records, wearables, genomics and drug discovery.[2-6] Healthcare AI has attracted huge financial investment, with interest from both the public and private sectors in recent years.[7] Notably, by the end of 2019, the 50 largest private sector investments into healthcare AI had reached a total of $8.5 billion and further growth in investment is anticipated, indicating the significant potential for AI to impact human health.[7,8]

"Surgical pathology" or "Histopathology" is a medical specialty where samples of tissue or cells are examined to provide vital information to clinicians and patients on the diagnosis, treatment and prognosis of cancers and other diseases.[9] Exciting AI innovations developed for digital pathology, usually derived from a technology called "whole slide imaging", have been applied to a range of diseases, including prostate, skin, breast, liver, colorectal and kidney pathologies.[10-15] Whole slide imaging involves the scanning and digitisation of a whole glass microscope slide to store it as a high resolution image that is available for the pathologist to view on a computer at any time.[16] Diagnostic reports, molecular techniques and other laboratory imaging modalities are also targets for AI research.[17-22] It is anticipated that these advancements will alter the way pathologists diagnose and influence patient management in future. However, as with other areas of healthcare, there are many barriers to achieving successful clinical implementation of these solutions and the introduction of any new technologies must be supported by robust evidence.[23,24]

The ability to comprehensively report research findings to make them understandable and usable by other researchers and clinicians is an essential skill for academics, and is key in ensuring broader scientific and technological progress. Unfortunately, there are multiple incentives and pressures for clinicians to rapidly publish research that does not necessarily include the essential components needed for both critical appraisal of a study and to make the work usable and useful to other researchers, and to the wider field.[25,26] Concerns around the quality of research methodology reported in scientific studies more generally first appeared in the literature in the early 20th century and it was recognised that this issue was impacting the quality of evidence and robustness of conclusions reached within research.[27] In an endeavour to overcome this, the first calls for reporting guidelines and attempts to produce them materialised in the 1980s and 1990s, and probably the most widely known early reporting guideline, the CONSORT statement for reporting of clinical trials, was published in 1996.[27,28] A reporting guideline is a tool or checklist used to guide the researcher in providing the minimum essential criteria needed when summarising their research study for publication.[29] A library of reporting guidelines for a range of study types can be found at the EQUATOR network website https://www.equator-network.org/reporting-guidelines/.[30] The EQUATOR network was established in 2008 as an international collaboration of academics and other stakeholders, with an aim to promote transparent and accurate reporting of medical research.[29] Further reporting guidelines have been developed for use in a variety of contexts since the initial CONSORT statement and these can be found within the EQUATOR network's online library.[29,30] Following decades of work to improve reporting quality, there is evidence to show that completeness of reporting improves with the use and endorsement of reporting guidelines.[31-36]

Concern that AI research is particularly vulnerable to issues of bias, methodological quality, and poor reporting, has been raised in several recent studies and risks leading to "research waste" following the huge investment into these technologies. A systematic review by Liu et al, in 2019 demonstrated that few of the deep learning studies examined provided externally validated results, few compared the performance of AI studies and healthcare professionals using the same sample and reporting of AI studies was frequently poor.[37] A systematic review of medical AI trials by Nagendran et al. in 2020 showed that most AI trials were at high risk of bias and deviated from established reporting standards.[38] Finally, a systematic review of machine learning for COVID-19 in radiology imaging by Roberts et al. in 2021 showed a high prevalence of deficiencies in the methodology and reporting, and of 320 relevant studies, none of these were of potential clinical use due to methodological flaws and underlying biases.[39] Wider awareness, endorsement and use of reporting guidelines in AI research is one method that can be used to start tackling these issues.

Few studies have addressed the issue of reporting in computational pathology. In previous work, our group published a study examining reporting of AI diagnostic accuracy studies in pathology conference abstracts which demonstrated that reporting was suboptimal, reporting guidance was not used or endorsed and that work was needed to address areas of potential bias in AI studies.[40] A study by Hogan et al. in 2020 examined reporting of diagnostic accuracy studies more generally in pathology journals and showed that better enforcement of reporting guideline use was needed, as incomplete reporting was prevalent.[41] Radiology shares some similarities with digital pathology, in terms of providing diagnoses and using medical imaging systems. A study by Dratsch et al. in 2020 showed similar issues with incomplete reporting and poor use and endorsement of reporting guidance in AI diagnostic accuracy conference abstracts.[42]

In recognition of these concerns, the intention of this review is to highlight the tools and checklists that are readily available to those working in pathology AI research. Many of these resources are quick and easy to incorporate into an existing research study and will help to avoid unintentional errors and omissions by authors. Greater awareness of these guidelines could increase uptake and improve the quality of research reporting in computational pathology.

**Materials and methods**

All records from the EQUATOR network library of reporting guidelines https://www.equator-network.org/reporting-guidelines/ were systematically searched up to and including 3rd August 2022. Records were screened against defined inclusion and exclusion criteria.[30] EQUATOR network library records were screened by one reviewer (CM) and where relevance was unclear, the full paper was reviewed and where required, discussed with a second reviewer (DT). Records were screened against the following criteria:

Inclusion

- Published reporting guideline on EQUATOR Network website
- Directly applicable to pathology AI research (i.e. study reported could have a primary focus on AI and any relevant study type is acceptable)
- Applicable to more than a single disease or pathological subspecialty.

Exclusion

- Guidance with only peripheral relevance to computational pathology research
- Education / training focused reporting guidelines
- Guidance covering only generic statistical analysis within a study
- Guidance focused on style, rather than content of reporting

The inclusion and exclusion criteria were chosen to provide the most relevant range of resources. Recent literature reviews of pathology AI were examined prior to commencing the study to understand current focuses of research and to determine the categories for required reporting guidelines.[24,43-46] Five main categories of guidelines were determined, with an additional sixth category containing resources useful for multiple stages and study types. Categories were as follows:

- Literature reviews & research priorities
- Discovery stage
- Clinical trial stage
- Implementation stage
- Post implementation & guideline development
- Multiple stages

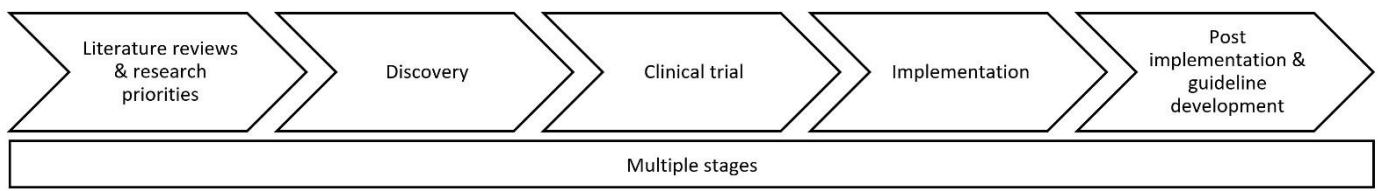

*Figure 1:* Stages of AI research with reporting guidelines available

Final guidelines for inclusion were summarised into tables by category and included information on the study type or topic, guideline name or description, the reference and direct website link for each resource. Guidelines listed as currently in development were reviewed separately for additional potential relevance to computational pathology research and it was noted that one record listed as "in development" had since been published online. This record was subsequently included within the results of published guidelines.

Ethical approval

This study examines previously published records and does not include any new human data or tissue that require ethical approval and consent. The authors assume that the studies examined were conducted after ethical approval and consent, and in accordance with the Declaration of Helsinki.

**Results**

A total of 530 records were identified and screened across a search that was completed in August 2022. The screening process is summarised in **Figure 2**. 72 relevant records were identified and are outlined across **Tables 1-6**. Relevant guidelines currently still in development were also noted in **Table 7**.

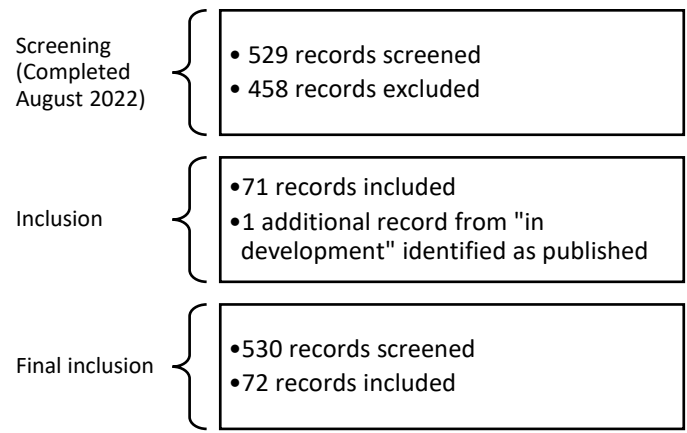

*Figure 2:* Screening of reporting guidelines

Literature review & research priorities

Fifteen resources were identified covering literature reviews and health research priorities and these are outlined in **Table 1**. This included guidelines and extensions for systematic reviews, guidance for scoping reviews, literature searches and systematic reviews without meta-analysis.

| LITERATURE REVIEW & RESEARCH PRIORITIES | | | |
|---|---|---|---|
| Study type / topic | Guideline | Reference | Link |
| Health research priorities | REPRISE (research priority setting studies) | Tong, A et al. 2019 [47] | https://www.equator-network.org/reporting-guidelines/reporting-guideline-for-priority-setting-of-health-research-reprise/ |
| Literature search | A structured approach to documenting a search strategy for publication: a 12 step guideline for authors | Kable, AK et al. 2012 [48] | https://www.equator-network.org/reporting-guidelines/a-structured-approach-to-documenting-a-search-strategy-for-publication-a-12-step-guideline-for-authors/ |
| | Reporting standards for literature searches and report inclusion criteria: making research syntheses more transparent and easy to replicate | Atkinson, KM et al. 2015 [49] | https://www.equator-network.org/reporting-guidelines/reporting-standards-for-literature-searches-and-report-inclusion-criteria-making-research-syntheses-more-transparent-and-easy-to-replicate/ |
| | "Brimful of STARLITE": toward standards for reporting literature searches | Booth, A. 2006 [50] | https://www.equator-network.org/reporting-guidelines/brimful-of-starlite-toward-standards-for-reporting-literature-searches/ |
| Scoping review | PRISMA-ScR | Tricco, AC et al. 2018 [51] | https://www.equator-network.org/reporting-guidelines/prisma-scr/ |
| Systematic review | Cochrane Handbook for Systematic Reviews of Interventions Version 6.1 | Higgins, JPT et al. 2020 [52] | https://www.equator-network.org/reporting-guidelines/cochrane-handbook-for-systematic-reviews-of-interventions/ |
| | Finding What Works in Health Care: Standards for Systematic Reviews. Chapter 5 – Standards for Reporting Systematic Reviews | Morton, S et al. 2011 [53] | https://www.equator-network.org/reporting-guidelines/finding-what-works-in-health-care-standards-for-systematic-reviews-chapter-5-standards-for-reporting-systematic-reviews/ |
| | PRISMA 2020 | Page, MJ et al. 2021 [54] | https://www.equator-network.org/reporting-guidelines/prisma/ |
| | PRISMA for Abstracts (review abstracts) | Contained within PRISMA 2020 statement paper: Page, MJ et al. 2021 [54] | https://www.equator-network.org/reporting-guidelines/prisma-abstracts/ |
| | PRISMA DTA (diagnostic test accuracy studies) | McInnes, MDF et al. 2018 [55] | https://www.equator-network.org/reporting-guidelines/prisma-dta/ |
| | PRISMA DTA for Abstracts (diagnostic test accuracy study review abstracts) | Cohen, JF et al. 2021 [56] | https://www.equator-network.org/reporting-guidelines/prisma-dta-for-abstracts/ |
| | PRISMA-P (protocols) | Moher, D et al. 2015 [57] | https://www.equator-network.org/reporting-guidelines/prisma-protocols/ |
| | PRISMA-S (literature searches) | Rethlefsen, ML et al 2021 [58] | https://www.equator-network.org/reporting-guidelines/prisma-s/ |
| | SWiM (without meta-analysis) | Campbell, M et al. 2020 [59] | https://www.equator-network.org/reporting-guidelines/synthesis-without-meta-analysis-swim-in-systematic-reviews-reporting-guideline/ |
| | Systematic Reviews. CRD's guidance for undertaking reviews in health care | Akers, J et al. 2009 [60] | https://www.equator-network.org/reporting-guidelines/systematic-reviews-crds-guidance-for-undertaking-reviews-in-health-care/ |

*Table 1: Reporting guidelines for literature reviews and research priorities*

Discovery stage

Thirteen published guidelines were identified as most relevant to the discovery stage of pathology AI research and these are outlined in **Table 2**. This included a range of study types and topics, including prediction models, diagnostic accuracy studies, pre-clinical research, patents, and image analysis challenges.

| | DISCOVERY | | |
|---|---|---|---|
| Study type / topic | Guideline | Reference | Link |
| AI in medical imaging | CLAIM | Mongan, J et al. 2020[61] | Still listed as in development, but published online: https://www.equator-network.org/library/reporting-guidelines-under-development/reporting-guidelines-under-development-for-other-study-designs/#CLAIM |
| Clinical AI modeling | MI-CLAIM | Norgeot, B et al. 2020[62] | https://www.equator-network.org/reporting-guidelines/minimum-information-about-clinical-artificial-intelligence-modeling-the-mi-claim-checklist/ |
| Databases | Reporting to Improve Reproducibility and Facilitate Validity Assessment for Healthcare Database Studies V1.0 | Wang, SV et al. 2017[63] | https://www.equator-network.org/reporting-guidelines/reporting-to-improve-reproducibility-and-facilitate-validity-assessment-for-healthcare-database-studies-v1-0/ |
| Diagnostic accuracy studies | STARD 2015 | Bossuyt, PM et al. 2015[64] | https://www.equator-network.org/reporting-guidelines/stard/ |
| | STARD for Abstracts | Cohen, J. F et al. 2017[65] | https://www.equator-network.org/reporting-guidelines/stard-abstracts/ |
| Image analysis Challenges | BIAS | Maier-Hein, L et al. 2020[66] | https://www.equator-network.org/reporting-guidelines/bias-transparent-reporting-of-biomedical-image-analysis-challenges/ |
| Patents | The Reporting Items for Patent Landscapes statement | Smith, JA et al. 2018[67] | https://www.equator-network.org/reporting-guidelines/the-reporting-items-for-patent-landscapes-statement/ |
| Pre-clinical research | A call for transparent reporting to optimize the predictive value of preclinical research | Landis, SC et al. 2012[68] | https://www.equator-network.org/reporting-guidelines/a-call-for-transparent-reporting-to-optimize-the-predictive-value-of-preclinical-research/ |
| | ARRIVE guidelines 2.0 | Percie du Sert, N et al. 2020[69] | https://www.equator-network.org/reporting-guidelines/improving-bioscience-research-reporting-the-arrive-guidelines-for-reporting-animal-research/ |
| Prediction models | Guidelines for Developing and Reporting Machine Learning Predictive Models in Biomedical Research: A Multidisciplinary View | Luo, W et al. 2016[70] | https://www.equator-network.org/reporting-guidelines/guidelines-for-developing-and-reporting-machine-learning-predictive-models-in-biomedical-research-a-multidisciplinary-view/ |
| | TRIPOD | Collins, GS et al. 2015[71] | https://www.equator-network.org/reporting-guidelines/tripod-statement/ |
| | TRIPOD for Abstracts | Heus, P et al. 2020[72] | https://www.equator-network.org/reporting-guidelines/tripod-for-abstracts/ |
| Reliability & agreement studies | GRRAS | Kottner, J et al. 2011[73] | https://www.equator-network.org/reporting-guidelines/guidelines-for-reporting-reliability-and-agreement-studies-grras-were-proposed/ |
| Telemedicine | Recommendations for the improved effectiveness and reporting of telemedicine programs in developing countries: results of a systematic literature review | Khanal, S et al. 2015[74] | https://www.equator-network.org/reporting-guidelines/recommendations-for-the-improved-effectiveness-and-reporting-of-telemedicine-programs-in-developing-countries-results-of-a-systematic-literature-review/ |

*Table 2: Reporting guidelines for discovery stage studies*

Clinical trial stage

Sixteen resources for the clinical trial stage of AI research are described in **Table 3**. This includes general guidance for clinical trials, along with guidance for digital technologies specifically.

| | CLINICAL TRIAL | | |
|---|---|---|---|
| Study type / topic | Guideline | Reference | Link |
| Clinical Trials (general) | CONSORT 2010 | Schulz, KF et al. 2010 [75] | https://www.equator-network.org/reporting-guidelines/consort/ |
| | SPIRIT 2013 (protocols) | Chan, AW et al. 2013 [76] | https://www.equator-network.org/reporting-guidelines/spirit-2013-statement-defining-standard-protocol-items-for-clinical-trials/ |
| Digital technologies | CONSORT-AI (clinical trials of AI) | Liu, X et al. 2020 [77] | https://www.equator-network.org/reporting-guidelines/consort-artificial-intelligence/ |
| | CONSORT-EHEALTH (eHealth clinical trials) | Eysenbach, G 2011 [78] | https://www.equator-network.org/reporting-guidelines/consort-ehealth-improving-and-standardizing-evaluation-reports-of-web-based-and-mobile-health-interventions/ |
| | STARE-HI (evaluation studies in health informatics) | Talmon, J et al. 2009 [79] | https://www.equator-network.org/reporting-guidelines/stare-hi-statement-on-reporting-of-evaluation-studies-in-health-informatics/ |
| | Systematic prioritization of the STARE-HI reporting items. An application to short conference papers on health informatics evaluation | de Keizer, NF et al. 2012 [80] | https://www.equator-network.org/reporting-guidelines/systematic-prioritization-of-the-stare-hi-reporting-items-an-application-to-short-conference-papers-on-health-informatics-evaluation/ |
| | SPIRIT-AI (clinical trial protocols for AI) | Rivera, SC et al. 2020 [81] | https://www.equator-network.org/reporting-guidelines/spirit-artificial-intelligence/ |
| | DECIDE-AI (early stage clinical evaluation of decision support systems) | Vasey, B et al. 2022 [82] | https://www.equator-network.org/reporting-guidelines/reporting-guideline-for-the-early-stage-clinical-evaluation-of-decision-support-systems-driven-by-artificial-intelligence-decide-ai/ |
| Economics for trials | Economic evaluation alongside randomised controlled trials: design, conduct, analysis, and reporting | Petrou, S et al. 2011 [83] | https://www.equator-network.org/reporting-guidelines/economic-evaluation-alongside-randomised-controlled-trials-design-conduct-analysis-and-reporting/ |
| | Good research practices for cost-effectiveness analysis alongside clinical trials: the ISPOR RCT-CEA Task Force report | Ramsey, S et al. 2015 [84] | https://www.equator-network.org/reporting-guidelines/good-research-practices-for-cost-effectiveness-analysis-alongside-clinical-trials-the-ispor-rct-cea-task-force-report/ |
| Outcomes | COS-STAP (outcome protocol items) | Kirkham, JJ et al. 2019 [85] | https://www.equator-network.org/reporting-guidelines/core-outcome-set-standardised-protocol-items-the-cos-stap-statement/ |
| | COS-STAR | Kirkham, JJ et al. 2016 [86] | https://www.equator-network.org/reporting-guidelines/cos-star-statement/ |
| Noninferiority & equivalence studies | Reporting of noninferiority and equivalence randomized trials: extension of the CONSORT 2010 statement | Piaggio, G et al. 2012 [87] | https://www.equator-network.org/reporting-guidelines/consort-non-inferiority/ |
| Non-randomised studies | Guidelines for reporting non-randomised studies | Reeves, BC et al 2004 [88] | https://www.equator-network.org/reporting-guidelines/guidelines-for-reporting-non-randomised-studies/ |
| Pathology | SPIRIT-Path | Kendall, TJ et al. 2021 [89] | https://www.equator-network.org/reporting-guidelines/spirit-path/ |
| Pilot & feasibility trials | CONSORT 2010 statement: extension to randomised pilot and feasibility trials | Eldridge, SM et al. 2016 [90] | https://www.equator-network.org/reporting-guidelines/consort-2010-statement-extension-to-randomised-pilot-and-feasibility-trials/ |

*Table 3: Reporting guidelines for clinical trial stage studies*

Implementation stage

Six guidelines most useful for the implementation stage are outlined in **Table 4**. There is published guidance available for performing implementation studies, health economics evaluations and stakeholder analysis.

| IMPLEMENTATION | | | |
|---|---|---|---|
| Study type / topic | Guideline | Reference | Link |
| Economics | CHEERS 2022 (health economic evaluations) | Husereau, D et al. 2022 [91] | https://www.equator-network.org/reporting-guidelines/cheers/ |
| | Health-Economic Analyses of Diagnostics: Guidance on Design and Reporting | van der Pol, S et al. 2021 [92] | https://www.equator-network.org/reporting-guidelines/health-economic-analyses-of-diagnostics-guidance-on-design-and-reporting/ |
| | Increasing the generalizability of economic evaluations: recommendations for the design, analysis, and reporting of studies | Drummond, M et al. 2005 [93] | https://www.equator-network.org/reporting-guidelines/increasing-the-generalizability-of-economic-evaluations-recommendations-for-the-design-analysis-and-reporting-of-studies/ |
| | Recommendations for Conduct, Methodological Practices, and Reporting of Cost-effectiveness Analyses: Second Panel on Cost-Effectiveness in Health and Medicine | Sanders, GD et al. 2016 [94] | https://www.equator-network.org/reporting-guidelines/recommendations-for-reporting-cost-effectiveness-analyses/ |
| Implementation studies | StaRI | Pinnock, H et al. 2017 [95] | https://www.equator-network.org/reporting-guidelines/stari-statement/ |
| Stakeholder analysis | Stakeholder analysis in health innovation planning processes: A systematic scoping review | Franco-Trigo, L et al. 2020 [96] | https://www.equator-network.org/reporting-guidelines/stakeholder-analysis-in-health-innovation-planning-processes-a-systematic-scoping-review/ |

*Table 4*: Reporting guidelines for implementation stage studies

Post-implementation and guideline development

**Table 5** contains resources focused on developing clinical guidelines and reporting quality improvement exercises. Eight guidelines were identified as being most useful in this category.

| POST-IMPLEMENTATION & GUIDELINES | | | |
|---|---|---|---|
| Study type / topic | Guideline | Reference | Link |
| Guidelines | AGREE (clinical practice guidelines) | Brouwers, MC et al. 2016 [97] | https://www.equator-network.org/reporting-guidelines/the-agree-reporting-checklist-a-tool-to-improve-reporting-of-clinical-practice-guidelines/ |
| | CheckUp (updated guidelines) | Vernooij, RWM et al. 2017 [98] | https://www.equator-network.org/reporting-guidelines/reporting-items-for-updated-clinical-guidelines-checkup/ |
| | How to write a guideline: a proposal for a manuscript template that supports the creation of trustworthy guidelines | Nieuwlaat R et al. 2021 [99] | https://www.equator-network.org/reporting-guidelines/how-to-write-a-guideline-a-proposal-for-a-manuscript-template-that-supports-the-creation-of-trustworthy-guidelines/ |
| | RIGHT (clinical practice guidelines) | Chen, Y et al. 2017 [100] | https://www.equator-network.org/reporting-guidelines/right-statement/ |
| | RIGHT-PVG (public versions of guidelines) | Wang, X et al. 2021 [101] | https://www.equator-network.org/reporting-guidelines/the-reporting-checklist-for-public-versions-of-guidelines-right-pvg/ |
| | Standardized reporting of clinical practice guidelines: a proposal from the Conference on Guideline Standardization | Shiffman, RN et al. 2003 [102] | https://www.equator-network.org/reporting-guidelines/standardized-reporting-of-clinical-practice-guidelines-a-proposal-from-the-conference-on-guideline-standardization/ |
| Quality Improvement | A new structure for quality improvement reports | Moss, F et al. 1999 [103] | https://www.equator-network.org/reporting-guidelines/a-new-structure-for-quality-improvement-reports/ |
| | SQUIRE 2.0 | Ogrinc, G et al. [104] | https://www.equator-network.org/reporting-guidelines/squire/ |

*Table 5:* Reporting guidelines for post-implementation and guideline development

Multiple stages

Thirteen published resources were identified that may be useful to researchers working on pathology AI at multiple stages towards clinical implementation. Qualitative research and patient and public involvement are increasingly important as the field moves closer to clinical use of these products. Handling bioresources, data linkage and images are may also be relevant to many pathology focused AI research study types. The details of these guidelines are outlined in **Table 6**.

| MULTIPLE STAGES | | | |
|---|---|---|---|
| **Study type / topic** | **Guideline** | **Reference** | **Link** |
| Bioresources | CoBRA (citation of bioresources) | Bravo, E et al. 2015 [105] | https://www.equator-network.org/reporting-guidelines/cobra/ |
| | BRISQ (biospecimen reporting) | Moore, HM et al. 2011 [106] | https://www.equator-network.org/reporting-guidelines/brisq/ |
| Data linkage | Development and validation of reporting guidelines for studies involving data linkage | Bohensky, MA et al 2011 [107] | https://www.equator-network.org/reporting-guidelines/development-and-validation-of-reporting-guidelines-for-studies-involving-data-linkage/ |
| Images | CLIP principles (clinical & laboratory images in publications) | Lang, TA et al. 2012 [108] | https://www.equator-network.org/reporting-guidelines/documenting-clinical-and-laboratory-images-in-publications-the-clip-principles/ |
| Patient & public involvement | GRIPP2 | Staniszewska, S et al. 2017 [109] | https://www.equator-network.org/reporting-guidelines/gripp2-reporting-checklists-tools-to-improve-reporting-of-patient-and-public-involvement-in-research/ |
| Qualitative research | COREQ (interviews & focus groups) | Tong, A et al. 2007 [110] | https://www.equator-network.org/reporting-guidelines/coreq/ |
| | Guidance for publishing qualitative research in informatics | Ancker, JS et al. 2021 [111] | https://www.equator-network.org/reporting-guidelines/guidance-for-publishing-qualitative-research-in-informatics/ |
| | Qualitative research: standards, challenges, and guidelines | Malterud, K 2001 [112] | https://www.equator-network.org/reporting-guidelines/qualitative-research-standards-challenges-and-guidelines/ |
| | Revealing the wood and the trees: reporting qualitative research | Blignault, I. et al. 2009 [113] | https://www.equator-network.org/reporting-guidelines/revealing-the-wood-and-the-trees-reporting-qualitative-research/ |
| | Standards for reporting qualitative research: a synthesis of recommendations | O'Brien, BC et al. 2014 [114] | https://www.equator-network.org/reporting-guidelines/srqr/ |
| Qualitative research: surveys | CHERRIES (web surveys) | Eysenbach, G 2004 [115] | https://www.equator-network.org/reporting-guidelines/improving-the-quality-of-web-surveys-the-checklist-for-reporting-results-of-internet-e-surveys-cherries/ |
| | CROSS | Sharma, A et al. 2021 [116] | https://www.equator-network.org/reporting-guidelines/a-consensus-based-checklist-for-reporting-of-survey-studies-cross/ |
| | Good practice in the conduct and reporting of survey research | Kelley, K et al. 2003 [117] | https://www.equator-network.org/reporting-guidelines/good-practice-in-the-conduct-and-reporting-of-survey-research/ |

*Table 6: Reporting guidelines relevant at multiple stages of pathology AI research*

Reporting guidelines in development

There are guidelines directly relevant to AI research that are currently in development. Four key guidelines are highlighted in **Table 7** and are anticipated to be available soon.

| | IN DEVELOPMENT | | |
|---|---|---|---|
| Study type / topic | Guideline | Protocol | Link |
| AI prediction models | TRIPOD-AI: Reporting of Artificial Intelligence and Machine Learning Studies | Collins, G et al. 2021[118] | All guidelines in development listed here: https://www.equator-network.org/library/reporting-guidelines-under-development/ |
| Diagnostic accuracy studies | STARD-AI: Reporting Guidelines for Diagnostic Accuracy Studies Evaluating Artificial Intelligence Interventions | Sounderajah, V et al. 2021[119] | |
| | STARD-AI for Abstracts: Reporting Guidelines for Diagnostic Accuracy Studies Evaluating Artificial Intelligence Intervention | | |
| Large-scale clinical informatics research | GLACIeR: Guidelines for Large-Scale, Applied Clinical Informatics Research | Not available | |

*Table 7:* AI reporting guidelines in development

**Discussion**

The intention of this review is to highlight the range of guidance for reporting research that is published and readily available to those working in computational pathology. Over 70 resources were identified as useful and outlined across a range of study types and topics. Frequently, these comprise simple checklists where components can be ticked off by the researcher during the course of writing an article. These are increasingly included as part of a journal's author guidelines, with some journals requiring completion of a checklist at the time of manuscript submission. Many guidelines are available to choose from and potential crossover between guidelines is a consideration, possibly making it more challenging to identify the most appropriate for use in a given study type. To address these concerns and given the issues described earlier with incomplete reporting, summarising, and categorising the range of resources available may be helpful to those conducting and reporting studies within this field.[40,41] It must also be stated that whilst usually it will be very clear which is the most appropriate guideline to follow, reviewing the options and selecting the most appropriate guidance to fit the context may be required by the researcher. A selection of key recommended guidance for each research stage described is outlined in **Table 8**. It is hoped that generating better awareness of these resources among researchers will benefit computational pathology overall and improve reporting quality.

| KEY GUIDANCE | | | |
|---|---|---|---|
| Guideline | Explanation | Reference | Link |
| *Literature review & research priorities* | | | |
| PRISMA & extensions | Guidance for systematic reviews with extensions for abstracts, protocols, literature searches, diagnostic test accuracy studies and scoping reviews. | Main guideline: Page, MJ et al. 2021 [54] | https://www.equator-network.org/reporting-guidelines/prisma/ |
| *Discovery stage* | | | |
| CLAIM | Guidance for studies of AI in medical imaging | Mongan, J et al. 2020 [61] | Listed "in development", but published online: https://www.equator-network.org/library/reporting-guidelines-under-development/reporting-guidelines-under-development-for-other-study-designs/#CLAIM |
| STARD & extensions | Guidance for studies of diagnostic test accuracy and an extension for abstracts. An extension for AI is currently in development. | Main guideline: Bossuyt, PM et al. 2015 [64] | https://www.equator-network.org/reporting-guidelines/stard/ |
| TRIPOD & extensions | Guidance for prediction model studies and an extension for abstracts. An extension for AI is currently in development. | Main guideline: Collins, GS et al. 2015 [71] | https://www.equator-network.org/reporting-guidelines/tripod-statement/ |
| *Clinical trial stage* | | | |
| CONSORT-AI | A guideline extension of CONSORT for clinical trials of AI models. | Liu, X et al. 2020 [77] | https://www.equator-network.org/reporting-guidelines/consort-artificial-intelligence/ |
| SPIRIT-AI | A guideline extension of SPIRIT for clinical trial protocols of AI models. | Rivera, SC et al. 2020 [81] | https://www.equator-network.org/reporting-guidelines/spirit-artificial-intelligence/ |
| DECIDE-AI | Guidance for early stage clinical evaluation of decision support systems. | Vasey, B et al. 2022 [82] | https://www.equator-network.org/reporting-guidelines/reporting-guideline-for-the-early-stage-clinical-evaluation-of-decision-support-systems-driven-by-artificial-intelligence-decide-ai/ |
| *Implementation stage* | | | |
| CHEERS | Guidance for health economic evaluations. | Husereau, D et al. 2022 [91] | https://www.equator-network.org/reporting-guidelines/cheers/ |
| StaRI | Guidance for implementation studies. | Pinnock, H et al. 2017 [95] | https://www.equator-network.org/reporting-guidelines/stari-statement/ |
| *Post implementation & guidelines* | | | |
| AGREE | Guidance for writing clinical practice guidelines | Brouwers, MC et al. 2016 [97] | https://www.equator-network.org/reporting-guidelines/the-agree-reporting-checklist-a-tool-to-improve-reporting-of-clinical-practice-guidelines/ |
| RIGHT | | Chen, Y et al. 2017 [100] | https://www.equator-network.org/reporting-guidelines/right-statement/ |

*Table 8: A selection of key guidance for pathology AI researchers*

Related work

Several review articles have previously highlighted the EQUATOR Network resources and reporting guidelines across healthcare research in general, including articles by Simera et al. across 2009-2010[25,29,120]. More recently, others have examined guidelines intended for artificial intelligence in healthcare, with publications by Ibrahim et al. and Shelmerdine et al. in 2021.[121,122] These reviews examine the recent proliferation of AI specific extensions to established reporting guidelines and explain where these are most appropriately applied. Additionally, they feature other useful resources such as PROBAST-ML (a risk of bias assessment tool that is currently in development and intended to support the TRIPOD-AI guideline extension) and MINIMAR, a non- EQUATOR reporting guideline published in 2020 for studies reporting the use of AI systems in healthcare.[123-125] AI specific extensions for reporting in surgical pathology clinical trials has been discussed by our group previously.[126] However, the authors are not aware of a comprehensive review of reporting guidelines applied generally to computational pathology research prior to this work.

This study

The start of a new research project usually includes a review of the relevant literature surrounding the research question and guidance for this is presented in **Table 1**. Publishing any review of the literature starts with a literature search and resources are available for conducting and summarising these searches.[48-50] Where a more extensive review is required, the PRISMA guidelines, PRISMA extensions and other tools are available to assist with systematic reviews, including writing a protocol, and performing a review of diagnostic accuracy studies.[52-55,57,58,60,65] The SWiM guidance is available for systematic reviews without a meta-analysis and PRISMA-ScR can be applied to scoping reviews.[51,59] Guidance also exists for those setting health research priorities with the REPRISE guideline.[47]

At this time, the majority of pathology AI publications describe new discoveries developed in a research environment, encompassing an assortment of study designs (**Table 2**). Where AI is used to determine a pathological diagnosis, the STARD and STARD for Abstracts guidance are recommended.[64,65] AI specific extensions to these guidelines are currently in development and should be applied in future for this context (**Table 7**).[119] Guidance for prediction models includes the TRIPOD guidelines and there is also a TRIPOD-AI extension in development for these studies.[118] Resources addressing pre-clinical research are available and include the ARRIVE guidelines.[68-72] For digital technologies at the discovery stage, MI-CLAIM is intended for clinical AI modeling studies, BIAS is for image analysis challenges and specific recommendations are in place for telemedicine studies.[62,66,74] Other useful resources at this stage include recommendations for database studies, patents and reliability and agreement studies.[63,67,73]

Clinical trials of AI models are essential before implementation for clinical use on patients can be considered. A selection of resources for clinical trials are outlined in **Table 3**, addressing clinical trials in general (such as the CONSORT and SPIRIT guidelines) and several specifically addressing digital technologies.[75-81] Of note, CONSORT-AI and SPIRIT-AI are AI specific extensions, covering clinical trials and clinical trial protocols respectively.[77,81] Additional recommendations for specific study types address Noninferiority and equivalence studies, non-randomised studies, SPIRIT-Path addressing pathology specific issues in clinical trials and pilot and feasibility studies.[87-90] Moreover, economic evaluations for trials and clinical trial outcomes are also addressed.[83-86]

Clinical implementation of AI is a critical step in seeing benefit for patients following extensive research investment. Whilst guidance in this area is currently lacking, it is likely to see further research as the technology progresses, and some limited guidance is available at this stage (**Table 4**). StaRI is a guideline available specifically for reporting implementation studies.[95] Other useful resources deal with reporting of stakeholder analyses and multiple documents address health economic evaluations.[91-94,96]

Guidance following implementation of AI products is unsurprisingly very limited at this time, given the context of the current state and progress of AI use in surgical pathology. The most relevant material at this stage (**Table 5**) applies to developing clinical guidelines and performing quality improvement exercises.[97-101,103,104] Whilst these will be important considerations when AI products are fully implemented, more work will be needed at this stage in future.

An assortment of resources with relevance to computational pathology researchers across a variety of contexts are also available. **Table 6** summarises guidelines applicable to multiple stages of research. There is increasing emphasis on patient and public involvement in AI research and qualitative research will be essential throughout the process of developing and implementing AI products successfully into any clinical setting. A series of guidelines have been published to address studies in these areas.[109-117] Additional tools most applicable to pathology research include CoBRA and BRISQ for bioresources, CLIP for images and there is a guideline available to address data linkage.[105-108]

Limitations

The scope of this study included articles from the EQUATOR network website only and whilst their coordinators run regular searches to identify new published guidance, it is acknowledged that some additional relevant material may not be included within this library.[25] This study was designed to encompass a wide range of useful resources for common study types to address incomplete reporting within pathology AI research, however this is not intended to be an exhaustive list and there may be additional context specific tools that are not addressed here. The EQUATOR library is deliberately open and inclusive of a range of guidelines. As such, there may be differences in the methods used in their development. However, the EQUATOR team have published recommendations for those wishing to develop new guidelines to try and ensure robustness in this process.[27,127]

**Conclusions**

A wealth of published resources and guidelines are readily available to the researchers to assist in complete reporting of their research. Incomplete reporting is prevalent in pathology and AI research, and reporting guidelines can help to improve reporting quality. These resources can be easily included in a range of study types and can help to avoid unintentional omissions and errors. Improved reporting is one method to tackle wider concerns around the quality of evidence underpinning new AI technologies intended for future clinical use.

**Acknowledgements**

Dr McGenity and Prof. Treanor are funded by National Pathology Imaging Co-operative (NPIC). NPIC (project no. 104687) is supported by a £50m investment from the Data to Early Diagnosis and Precision Medicine strand of the Government's Industrial Strategy Challenge Fund, managed and delivered by UK Research and Innovation (UKRI). Dr McGenity is funded by Leeds Hospitals Charity and the National Institute for Health Research (NIHR).

**Author contributions**

D.T. and C.M. conceived and designed the study. C.M. screened records and compiled results with assistance from DT. C.M. and D.T. drafted and revised the manuscript. C.M. and D.T. read and approved the final version of the manuscript.

**Competing interests**

The authors have no competing interests to declare.